%% file: QConfinement.tex
\input{aipcheck}

\documentclass[
    ,final            
    ,numberedheadings 
  ]
  {aipproc}

\layoutstyle{6x9}

\usepackage{amstext}
\usepackage{amsfonts}
\usepackage{amssymb}
\usepackage{color}
\usepackage{epsfig}

\begin{document}

\title{Properties of confining gauge field configurations in the pseudoparticle approach}

\classification{11.15.-q.}

\keywords{SU(2) Yang-Mills theory, pseudoparticles, confinement.}

\author{Marc Wagner}{
  address={Institute~for~Theoretical~Physics~III, University~of~Erlangen-N{\"u}rnberg, Staudtstra{\ss}e~7, 91058~Erlangen, Germany}
}

\begin{abstract}
The pseudoparticle approach is a numerical method to approximate path integrals in SU(2) Yang-Mills theory. Path integrals are computed by summing over all gauge field configurations, which can be represented by a linear superposition of a small number of pseudoparticles with amplitudes and color orientations as degrees of freedom. By comparing different pseudoparticle ensembles we determine properties of confining gauge field configurations. Our results indicate the importance of long range interactions between pseudoparticles and of non trivial topological properties.
\end{abstract}

\maketitle



\section{The basic principle of the pseudoparticle approach}

The pseudoparticle approach \cite{Wagner:2005vs,Wagner:2006,Wagner:2006qn} is a numerical technique to approximate Euclidean path integrals. In this work we consider pure SU(2) Yang-Mills theory, where the expectation value of a quantity $\mathcal{O}$ is given by
\begin{eqnarray}
\label{EQN_01} & & \hspace{-0.7cm} \Big\langle \mathcal{O} \Big\rangle \ \ = \ \ \frac{1}{Z} \int DA \, \mathcal{O}[A] e^{-S[A]} \quad , \quad S[A] \ \ = \ \ \frac{1}{4 g^2} \int d^4x \, F_{\mu \nu}^a F_{\mu \nu}^a
\end{eqnarray}
with $F_{\mu \nu}^a = \partial_\mu A_\nu^a - \partial_\nu A_\mu^a + \epsilon^{a b c} A_\mu^b A_\nu^c$. Furthermore, the pseudoparticle approach is a tool to analyze the importance of certain classes of gauge field configurations with respect to confinement (c.f.\ section~\ref{SEC_01}).

The basic idea of the pseudoparticle approach is to consider only those gauge field configurations, which can be written as a sum of a fixed number ($\approx 400$) of pseudoparticles:
\begin{eqnarray}
\nonumber & & \hspace{-0.7cm} A_\mu^a(x) \ \ = \ \ \sum_i \mathcal{A}(i) \mathcal{C}^{a b}(i) a_{\mu,\textrm{inst.}}^b(x-z(i)) + \sum_j \mathcal{A}(j) \mathcal{C}^{a b}(j) a_{\mu,\textrm{antiinst.}}^b(x-z(j)) + \\
\label{EQN_02} & & \hspace{0.675cm} \sum_k \mathcal{A}(k) \mathcal{C}^{a b}(k) a_{\mu,\textrm{akyron}}^b(x-z(k))
\end{eqnarray}
with amplitudes $\mathcal{A}(i) \in \mathbb{R}$, color orientation matrices $\mathcal{C}^{a b}(i) \in \textrm{SO(3)}$ and positions $z(i) \in \mathbb{R}^4$ as degrees of freedom. Our standard choice of pseudoparticles are ``regular gauge instantons'', ``regular gauge antiinstantons'' and akyrons:
\begin{eqnarray}
\nonumber & & \hspace{-0.7cm} a_{\mu,\textrm{inst.}}^b(x) \ \ = \ \ \eta_{\mu \nu}^b x_\nu \frac{1}{x^2 + \lambda^2} \quad , \quad a_{\mu,\textrm{antiinst.}}^b(x) \ \ = \ \ \bar{\eta}_{\mu \nu}^b x_\nu \frac{1}{x^2 + \lambda^2} \quad , \\
\label{EQN_03} & & \hspace{0.675cm} a_{\mu,\textrm{akyron}}^b(x) \ \ = \ \ \delta^{b 1} x_\mu \frac{1}{x^2 + \lambda^2}
\end{eqnarray}
with $\eta_{\mu \nu}^b = \epsilon_{b \mu \nu} + \delta_{b \mu} \delta_{0 \nu} - \delta_{b \nu} \delta_{0 \mu}$ and $\bar{\eta}_{\mu \nu}^b = \epsilon_{b \mu \nu} - \delta_{b \mu} \delta_{0 \nu} + \delta_{b \nu} \delta_{0 \mu}$. Note, however, that instead of these pseudoparticles any set of localized gauge field configurations can be used (an example, Gaussian localized pseudoparticles, are discussed in section~\ref{SEC_01}). The path integral (\ref{EQN_01}) is approximated by an integration over amplitudes and color orientation matrices:
\begin{eqnarray}
\label{EQN_04} \Big\langle \mathcal{O} \Big\rangle \ \ = \ \ \frac{1}{Z} \int \left(\prod_i d\mathcal{A}(i) \, d\mathcal{C}(i)\right) \mathcal{O}(\mathcal{A}(i),\mathcal{C}(i)) e^{-S(\mathcal{A}(i),\mathcal{C}(i))} .
\end{eqnarray}

To be more concrete, we put $400$ pseudoparticles (pseudoparticle size $\lambda = 0.5$) with randomly chosen positions $z(i)$ inside a hyperspherical spacetime region with radius $3.0$ (this amounts to a pseudoparticle density of $1.0$) and compute the multidimensional integrals (\ref{EQN_04}) via Monte-Carlo sampling. Boundary effects are excluded by ``measuring'' observables sufficiently far away from the boundary.

\begin{figure}[b]
\input{FIG_001.pstex_t}
\caption{\label{FIG_01}The static quark antiquark potential plotted against the separation. \textbf{a)}~Standard choice of pseudoparticles (the data points have been fitted with $V_{\textrm{q} \bar{\textrm{q}}}(R) = V_0 - \alpha / R + \sigma R$). \textbf{b)}~Gaussian localized pseudoparticles of different size. \textbf{c)}~Akyron ensemble, instanton ensemble and mixed ensemble.}
\end{figure}
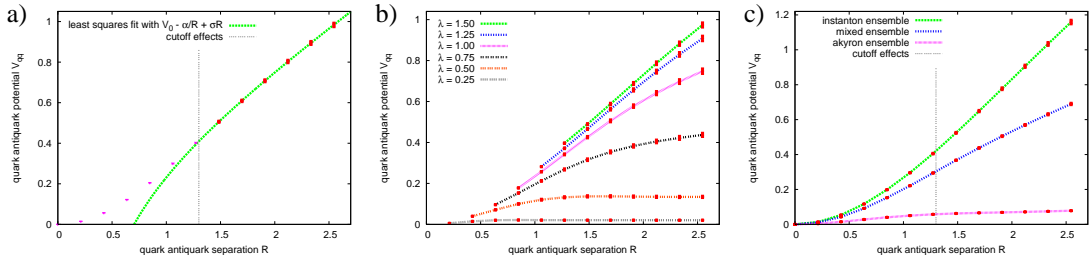

In \cite{Wagner:2006,Wagner:2006qn} it has been shown in detail that the pseudoparticle approach applied with around $400$ instantons, antiinstantons and akyrons is able to reproduce many essential features of SU(2) Yang-Mills theory. For example the static quark antiquark potential is linear for large separations (c.f.\ Figure~\ref{FIG_01}a), i.e.\ there is confinement. Like in lattice gauge theory the string tension $\sigma$ is an increasing function of the coupling constant $g$. Therefore, when the scale is set by identifying the numerical value of $\sigma$ with its physical value, e.g.\ $\sigma = 4.2 / \textrm{fm}^2$, the physical size of the hyperspherical spacetime region can be adjusted by choosing appropriate values for $g$. Furthermore, the string tension $\sigma$, the topological susceptibility $\chi$ and the critical temperature of the confinement deconfinement phase transition $T_\textrm{critical}$ scale consistently with $g$, i.e.\ the dimensionless ratios $\chi^{1/4} / \sigma^{1/2}$ and $T_\textrm{critical} / \sigma^{1/2}$ are constant with respect to $g$. They are also of the right order of magnitude when compared to lattice results.

Note that there are significant differences between our method and well-established instanton gas and instanton liquid models \cite{Schafer:1996wv}: (i) in general, our gauge field configurations (\ref{EQN_02}) are not even close to classical solutions, i.e.\ we do not consider a semiclassical limit, but try to approximate full quantum physics; (ii) the latter is related to the fact that our ``regular gauge pseudoparticles'' (\ref{EQN_03}) interact over large distances; this is so, because their gauge fields decrease like $1/|x|$ for large $|x|$ in contrast to the $1/|x|^3$-behavior of singular gauge instantons and antiinstantons; as we will demonstrate in section~\ref{SEC_01}, these long range interactions are intimately connected to confinement; (iii) because we include amplitudes as degrees of freedom, it is also possible to model small quantum fluctuations; furthermore, in the limit of infinitely many pseudoparticles any gauge field configuration can be represented according to (\ref{EQN_02}), i.e.\ in this limit the pseudoparticle approach is identical to full SU(2) Yang-Mills theory (c.f.\ \cite{Wagner:2006}).


\section{\label{SEC_01}Properties of confining gauge field configurations}

In the following we apply the pseudoparticle approach with different types of pseudoparticles to study the effect of different classes of gauge field configurations on confinement.


\paragraph*{Pseudoparticles of different size}

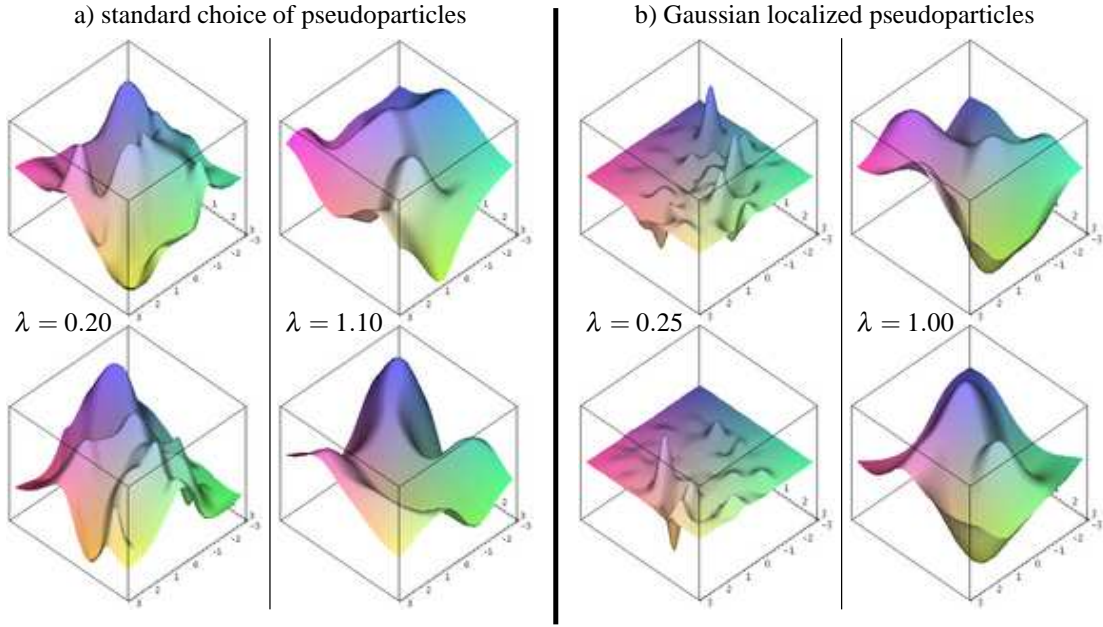
\begin{figure}[h]
\input{FIG_002.pstex_t}
\caption{\label{FIG_02}Typical gauge field configurations: one of the gauge field components $A_\mu^a$ plotted against two of the spacetime coordinates $x_\mu$ and $x_\nu$.}
\end{figure}

We have compared ensembles with different pseudoparticle size $\lambda$. Note that $\lambda$ strongly affects the shape of a pseudoparticle near its center, but has essentially no effect on the $1/|x|$ long range behavior (c.f.\ (\ref{EQN_03})). For $\lambda = 0.2 , \ldots , 1.1$ the static quark antiquark potential is essentially unaffected by the pseudoparticle size $\lambda$. This indicates that confinement is a consequence of the $1 / |x|$ long range behavior of the pseudoparticles, which is the same for all values of $\lambda$. Typical gauge field configurations for $\lambda = 0.2$ and for $\lambda = 1.1$ are shown in Figure~\ref{FIG_02}a. For both values of $\lambda$ the global structure is the same, but for $\lambda = 0.2$ there are additional local ultraviolet fluctuations. Apparently, these ultraviolet fluctuations have no effect on confinement and the string tension.


\paragraph*{Pseudoparticles with a limited range of interaction}

We have compared ensembles of Gaussian localized pseudoparticles with different size $\lambda$. These Gaussian localized pseudoparticles arise by replacing $1 / (x^2 + \lambda^2)$ in (\ref{EQN_03}) by $(1/\lambda^2) \exp(-x^2/2 \lambda^2)$. They have a limited range of interaction, which is proportional to $\lambda$. For short range pseudoparticles ($\lambda \leq 0.50$) there is little overlap between neighboring pseudoparticles and no confinement. For long range pseudoparticles ($\lambda \geq 1.00$) there is significant overlap and the static quark antiquark potential is confining (c.f.\ Figure~\ref{FIG_01}b). To put it another way, there is a close relation between ``pseudoparticle percolation'' and confinement. Typical gauge field configurations for $\lambda = 0.25$ and for $\lambda = 1.00$ are shown in Figure~\ref{FIG_02}b. For $\lambda = 0.25$ (no confinement) there are only local ultraviolet fluctuations, while for $\lambda = 1.00$ (confinement) there are global excitations. It seems that gauge field configurations responsible for confinement contain extended structures and large area excitations.


\paragraph*{Pseudoparticles without topological charge}

We have compared a pure akyron ensemble (400 akyrons), a pure instanton ensemble (200 instantons, 200 antiinstantons) and a ``mixed ensemble'' (150 instantons, 150 antiinstantons, 100 akyrons). There is no confinement in the akyron ensemble (c.f.\ Figure~\ref{FIG_01}c). That demonstrates that akyrons alone are not suited to reproduce correct Yang-Mills physics. Since superpositions of akyrons always have vanishing topological charge density (c.f.\ \cite{Wagner:2006qn}), it also supports the common expectation that confinement and topological charge are closely related. Both in the instanton and in the mixed ensemble there is confinement. However, comparing physically meaningful quantities in the instanton ensemble and in the mixed ensemble (e.g.\ $(\chi^{1/4} / \sigma^{1/2})_\textrm{instanton} = 0.26$, $(\chi^{1/4} / \sigma^{1/2})_\textrm{mixed} = 0.35$) with results from lattice calculations ($(\chi^{1/4} / \sigma^{1/2})_\textrm{lattice} = 0.49$ \cite{Teper:1998kw}) indicates that using akyrons is beneficial with respect to quantitative results.


\paragraph{Summary}

We have presented evidence that confining gauge field configurations contain extended structures and large area excitations. Topological charge also seems to play an important role. In contrast to that, confinement is not affected by ultraviolet fluctuations.


\begin{theacknowledgments}
I would like to thank M.~Faber, J.~Greensite and M.~Polikarpov for inviting me to the ``Quark Confinement and the Hadron Spectrum VII''-conference to give this talk, as well as J.~E.~Ribeiro for his support. Furthermore, it is a pleasure to thank F.~Lenz for many interesting and fruitful discussions.
\end{theacknowledgments}



\end{document}

%% file: FIG_001.pstex_t
\begin{picture}(0,0)%
\includegraphics{FIG_001.pstex}%
\end{picture}%
\setlength{\unitlength}{4144sp}%
\begingroup\makeatletter\ifx\SetFigFont\undefined%
\gdef\SetFigFont#1#2#3#4#5{%
  \reset@font\fontsize{#1}{#2pt}%
  \fontfamily{#3}\fontseries{#4}\fontshape{#5}%
  \selectfont}%
\fi\endgroup%
\begin{picture}(6570,1530)(1,-691)
\put(  1,704){\makebox(0,0)[lb]{\smash{{\SetFigFont{10}{12.0}{\familydefault}{\mddefault}{\updefault}{\color[rgb]{0,0,0}a)}%
}}}}
\put(2206,704){\makebox(0,0)[lb]{\smash{{\SetFigFont{10}{12.0}{\familydefault}{\mddefault}{\updefault}{\color[rgb]{0,0,0}b)}%
}}}}
\put(4411,704){\makebox(0,0)[lb]{\smash{{\SetFigFont{10}{12.0}{\familydefault}{\mddefault}{\updefault}{\color[rgb]{0,0,0}c)}%
}}}}
\end{picture}%

%% file: FIG_002.pstex_t
\begin{picture}(0,0)%
\includegraphics{FIG_002.pstex}%
\end{picture}%
\setlength{\unitlength}{4144sp}%
\begingroup\makeatletter\ifx\SetFigFont\undefined%
\gdef\SetFigFont#1#2#3#4#5{%
  \reset@font\fontsize{#1}{#2pt}%
  \fontfamily{#3}\fontseries{#4}\fontshape{#5}%
  \selectfont}%
\fi\endgroup%
\begin{picture}(6570,3765)(1,-2794)
\put(4951,839){\makebox(0,0)[b]{\smash{{\SetFigFont{10}{12.0}{\familydefault}{\mddefault}{\updefault}{\color[rgb]{0,0,0}b) Gaussian localized pseudoparticles}%
}}}}
\put(1576,839){\makebox(0,0)[b]{\smash{{\SetFigFont{10}{12.0}{\familydefault}{\mddefault}{\updefault}{\color[rgb]{0,0,0}a) standard choice of pseudoparticles}%
}}}}
\put(1666,-1006){\makebox(0,0)[lb]{\smash{{\SetFigFont{10}{12.0}{\familydefault}{\mddefault}{\updefault}{\color[rgb]{0,0,0}$\lambda=1.10$}%
}}}}
\put(3466,-1006){\makebox(0,0)[lb]{\smash{{\SetFigFont{10}{12.0}{\familydefault}{\mddefault}{\updefault}{\color[rgb]{0,0,0}$\lambda=0.25$}%
}}}}
\put(5086,-1006){\makebox(0,0)[lb]{\smash{{\SetFigFont{10}{12.0}{\familydefault}{\mddefault}{\updefault}{\color[rgb]{0,0,0}$\lambda=1.00$}%
}}}}
\put( 46,-1006){\makebox(0,0)[lb]{\smash{{\SetFigFont{10}{12.0}{\familydefault}{\mddefault}{\updefault}{\color[rgb]{0,0,0}$\lambda=0.20$}%
}}}}
\end{picture}%

%% file: QConfinement.bbl
\begin{thebibliography}{9}

\bibitem{Wagner:2005vs}
  M.~Wagner and F.~Lenz,
  PoS {\bf LAT2005} (2005) 315
  [arXiv:hep-lat/0510083].

\bibitem{Wagner:2006}
  M.~Wagner,
  PhD thesis, Institute for Theoretical Physics III, University of Erlangen-N\"urnberg (2006).

\bibitem{Wagner:2006qn}
  M.~Wagner,
  arXiv:hep-ph/0608090.

\bibitem{Schafer:1996wv}
  T.~Sch\"afer and E.~V.~Shuryak,
  Rev.\ Mod.\ Phys.\  {\bf 70} (1998) 323
  [arXiv:hep-ph/9610451].

\bibitem{Teper:1998kw}
  M.~J.~Teper,
  arXiv:hep-th/9812187.

\end{thebibliography}
